\documentclass[aps, prl,twocolumn,showpacs,floatfix,superscriptaddress,nofootinbib]{revtex4-2}
\usepackage{graphics}
\usepackage{xcolor}
\usepackage{amsmath, amssymb,amsthm}
\usepackage{mathtools}
\usepackage{braket}
\usepackage{bbold}
\usepackage[normalem]{ulem}
\usepackage[colorlinks=true,citecolor=blue,linkcolor=magenta]{hyperref}
\usepackage[caption=false]{subfig}

\newcommand{\diff}[1]{\mathrm{d}{#1}\,}
\renewcommand{\dfrac}[2]{\frac{\mathrm{d}{#1}}{\mathrm{d}{#2}}}
\newcommand{\pdfrac}[2]{\frac{\partial{#1}}{\partial{#2}}}

\newcommand{\bs}[1]{{\boldsymbol{#1}}}

\definecolor{MyBlue}{HTML}{1f77b4}
\definecolor{MyOrange}{HTML}{ff7f0e}
\definecolor{MyGreen}{HTML}{2ca02c}
\definecolor{MyRed}{HTML}{d62728}

\newcommand{\msout}[1]{\text{\sout{\ensuremath{#1}}}}

\newcommand{\mold}[1]{\textcolot{blue}{\msout}}

\begin{document}

\title{Quantum boomerang effect in systems without time reversal symmetry}

\author{Jakub Janarek}\email{jakub.janarek@uj.edu.pl}
\affiliation{Laboratoire Kastler Brossel, Sorbonne Universit\'e, CNRS,
ENS-PSL Research University, Coll\`ege de France, 4 Place Jussieu, 75005
Paris, France}
\affiliation{Instytut Fizyki Teoretycznej,
Uniwersytet Jagiello\'nski,  \L{}ojasiewicza 11, PL-30-348 Krak\'ow, Poland}

	\author{Beno\^it Gr\'emaud}\email{benoit.gremaud@cpt.univ-mrs.fr}
\affiliation{Aix-Marseille Universit\'e, Universit\'e de Toulon, CNRS, CPT, Marseille, France}



\author{Jakub Zakrzewski}\email{jakub.zakrzewski@uj.edu.pl}
\affiliation{Instytut Fizyki Teoretycznej,
Uniwersytet Jagiello\'nski,  \L{}ojasiewicza 11, PL-30-348 Krak\'ow, Poland}
\affiliation{Mark Kac Complex Systems Research Center, Uniwersytet Jagiello{\'n}ski, Krak{\'o}w, Poland}

\author{Dominique Delande}\email{dominique.delande@lkb.upmc.fr}
\affiliation{Laboratoire Kastler Brossel, Sorbonne Universit\'e, CNRS,
ENS-PSL Research University, Coll\`ege de France, 4 Place Jussieu, 75005
Paris, France}

\date{\today}

\begin{abstract}
In an Anderson localized system, a quantum particle with a nonzero initial velocity returns, on average, to its origin. This recently discovered behavior is known as the quantum boomerang effect. Time reversal invariance was initially thought to be a necessary condition for the existence of this phenomenon.
We theoretically analyze the impact of the symmetry breaking on the phenomenon using a one-dimensional system with a spin-orbit coupling and show that the  time reversal invariance is not necessary for the boomerang effect to occur. {We explain this behaviour giving  sufficient symmetry conditions for the boomerang effect to occur when time-reversal symmetry is broken.}
\end{abstract}
\date{\today}

\maketitle


\paragraph*{Introduction--}
Anderson localization (AL), the inhibition of transport due to the destructive interference of partial waves \cite{Anderson1958}, is one of the most important phenomena in disordered systems. AL was successfully observed in quantum systems \cite{Roati2008, Billy2008, Jendrzejewski2012, Manai2015, Semeghini2015}, as well as for acoustic~\cite{Hu2008} and electromagnetic waves~\cite{Chabanov2000, Schwartz2007}. 
While several manifestations of AL were discussed over the years,
 an entirely new phenomenon  was recently discovered -- the quantum boomerang effect
(QBE)~\cite{Prat2019}: The center of mass (CoM) of a quantum particle launched with a nonzero velocity in a disordered potential returns, on average, to its initial position when AL is present. In stark contrast, a classical particle will end, on average, at a finite distance (one transport mean free path) from its initial position.  
The phenomenon appears as a smoking gun of AL and occurs  in one- and higher-dimensional systems~\cite{Prat2019}, including pseudo-random potentials and the localization in momentum space as exhibited by a kicked rotor~\cite{Tessieri2021}. Very recently, an experimental observation of the QBE was reported for the kicked rotor~\cite{Sajjad2021}.
\begin{figure}
 	\includegraphics{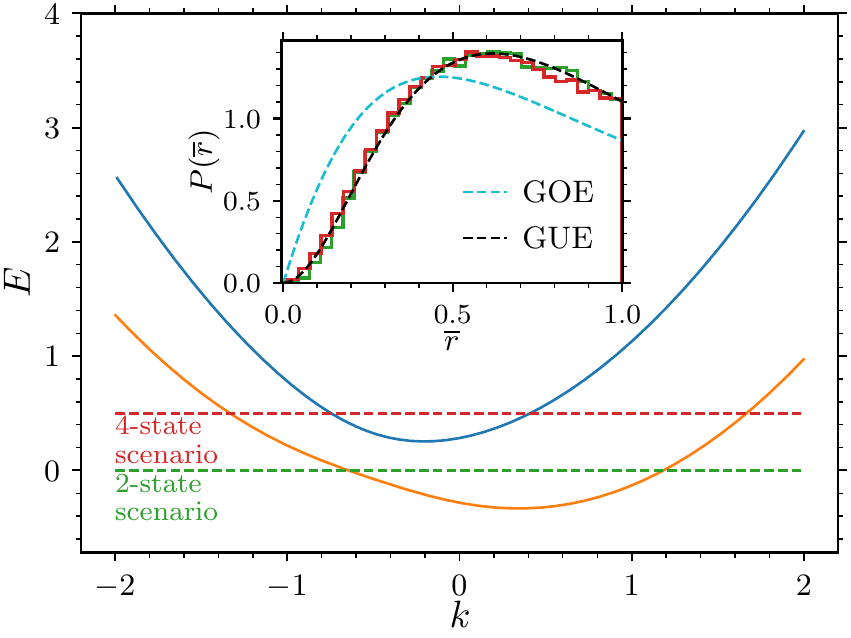}
	\caption{Spectrum of the Hamiltonian~(\ref{eq:Hamiltonian_0}) computed for $\gamma = 0.4$ and $\delta=\Omega=0.4$. Depending on the choice of the parameters, there are up to 4 possible eigenstates at a given energy with different velocities. The green (resp. red) dashed line represents an example of a 2-state (resp. 4-state) scenario. The inset displays the distributions of gap ratio $P(\bar{r})$ calculated around $E=0$ and $E=0.5$. The dashed lines are theoretical predictions for the GOE and GUE symmetry classes.
 	\label{fig:spectrum_and_rbar}
    }
 \end{figure}

Consider a one-dimensional Hamiltonian $H=p^2/2m+V(x)$, where $V(x)$ is a disordered potential, e.g. a Gaussian uncorrelated disorder~\cite{Prat2019}. For a wave packet with some initial velocity $\psi_0(x)=\mathcal{N}\exp(-x^2/2\sigma^2 + ik_0 x)$, the temporal evolution of the {CoM} is computed using $\langle x(t)\rangle=\int x\,\overline{|\psi(x,t)|^2}\diff{x}$, where $\overline{(\cdots)}$ denotes the average over disorder realizations.  The QBE assures that the 	
{CoM} returns to the origin, $\langle x(t\!=\!\infty)\rangle=0$. Until now, the existence of the QBE has been supported by time reversal invariance (TRI) arguments. In our work, we study the QBE in a system which breaks TRI. We show that the QBE may exist in such a situation. {First we show it on a simple example, both numerically and using the perturbative Berezinskii expansion. Later we formulate the sufficient conditions for QBE to occur when TRI is broken.}

 \paragraph*{The model--}
We consider a one-dimensional single-particle system with spin-orbit (SO) coupling {and Zeeman splitting} as the minimum ingredient to break TRI and all anti-unitary symmetries. For this purpose, we use the following well-known Hamiltonian \cite{Lin2011, Hamner2015}:
\begin{equation}\label{eq:Hamiltonian_0}
  H_0 = \frac{\hat{p}^2}{2m} + \gamma \hat{p}\sigma_z + \frac{\hbar\delta}{2}\sigma_z + \frac{\hbar \Omega}{2}\sigma_x,
\end{equation}
where  $\sigma_i$ are the standard Pauli matrices. The Hilbert space is spanned by 2-component complex-valued spinors. A specific experimental realization has been presented in~\cite{Lin2011, Hamner2015}, using a Raman coupling between two atomic states. $\gamma$  is the strength of the SO coupling, $\Omega$ the Rabi frequency of the Raman coupling and $\delta$ its detuning. {Due to translational invariance of $H_0$ the eigenstates can be labeled by wavenumbers $k$.} The spectrum of the Hamiltonian $H_0$ consists of two bands, $E_\pm(k) = \hbar^2 k^2/2m \pm \hbar/2\sqrt{(2\gamma k + \delta)^2 + \Omega^2},$ shown in Fig.~\ref{fig:spectrum_and_rbar}. In the numerical simulations, we assume $\hbar\!=\!1$ and $m\!=\!1$. {Every dimensional quantity is expressed through a chosen unit of length $a$, i.e. energies are in units of $\hbar m^{-1} a^{-2}$, times are in units of $\hbar^{-1} m a^2$, disorder strength $\eta$ in units of $\hbar^2 m^{-2} a^{-3}$, velocities  in units of $\hbar m^{-1} a^{-1}$, wave numbers in units of $a^{-1}$, etc.}

To study the QBE, we add to the Hamiltonian a disordered potential, $H=H_0 + V(x)$, similarly to \cite{Yue2020a}. For the disorder, we choose a Gaussian uncorrelated disorder:  $\overline{V(x)} = 0$, $\overline{V(x)V(x')}=\eta \delta(x-x')$, where $\eta$ denotes the disorder strength. The disorder is the same for both spin components.

\paragraph*{Condition for QBE--} Can we observe the QBE for our model? In~\cite{Prat2019}, it was proved that time reversal invariance of the Hamiltonian was a sufficient condition for the QBE to exist \cite{foot}.
 
For a spin-1/2 particle, the standard anti-unitary time reversal operator is $T=i\sigma_y K$, where $K$ denotes the the time reversal operator for a spinless particle, that is complex conjugation in configuration space~\cite{Haake2018} such that $x\!\to\!-x,t\!\to\!-t,p\!\to\!-p,K\psi(x)=\psi^*(x).$
If $\gamma\neq 0,$ the disordered Hamiltonian $H$ breaks time reversal invariance $T H T^{-1}\neq H$. 
This is however not the end of the story. As discussed in~\cite{Tessieri2021,Sajjad2021}, there are cases -- such as the kicked rotor -- where another anti-unitary symmetry, e.g. the product {$TP$} of the conventional time reversal{, $T$,} with spatial parity, $P$, such that $x\!\to\!-x,t\!\to\!t,p\!\to\!-p$, is sufficient to imply the QBE.
 
In our model, we have found that, for $\delta=0$, the Hamiltonian is invariant under the generalized anti-unitary operator $\mathcal{T}=i\sigma_z K.$ The disorder $V(x)$
breaks all the possible spatial symmetries. As a consequence, when $\delta,\Omega,\gamma$ are all non-zero, any generalized time reversal symmetry is broken, and the Hamiltonian belongs to the unitary symmetry class {represented by Gaussian Unitary Ensemble} (GUE) { (instead of the Gaussian Orthogonal Ensemble (GOE) appropriate for $PT$ or $\mathcal{T}$ symmetric case)}~\cite{Haake2018}. As a numerical evidence, the inset in Fig.~\ref{fig:spectrum_and_rbar} displays distributions of gap ratio $\bar{r}=\min(\delta_n/\delta_{n-1}, \delta_{n-1}/\delta_{n})$, where $\delta_n$ is the spacing between neighboring energy levels~\cite{Oganesyan2007,Atas2013a}. The figure shows a very good agreement of numerical data with the theoretical prediction for the GUE~\cite{Atas2013a}, proving that all anti-unitary symmetries are broken.

Weak disorder couples disorder-free eigenstates with the same energy, but different momenta and spin states. We first study the 2-state scenario, see Fig.~\ref{fig:spectrum_and_rbar}. 
{There are 2 momenta $k_{\pm}$ and 2 corresponding spinors $\ket{\chi_\pm}$ such that $\ket{k_{\pm}}\otimes\ket{\chi_\pm}$ are eigenstates of the disorder-free Hamiltonian~(\ref{eq:Hamiltonian_0}) with the same energy $E_0.$  Note that $k_-\!\neq\!-k_+,$ that the associated velocities $v_{\pm} = \frac{1}{\hbar}\frac{dE(k_\pm)}{dk}$ are not opposite and that the spin states $\ket{\chi_\pm}$ are not orthogonal.}

For the initial wavepacket, we {use either one of} the two states $\ket{\Psi^\pm_0}=\ket{\psi^\pm_0}\otimes\ket{\chi_\pm}$, where $\psi^\pm_0(x) = \mathcal{N}\exp(-x^2/2\sigma^2 + ik_\pm x)$, $\mathcal{N}$ is a normalization constant. {We distinguish results calculated for the two initial states by denoting them $\langle x_\pm\rangle$.} In our simulations, we have used $\sigma=50.$ The states $\ket{\Psi^\pm_0}$ are not exact eigenstates of $H_0$, they are approximations to monochromatic waves. However, based on the results of~\cite{Prat2019,Prat2017a}, we can safely assume that, as long as the initial state does not contain too many momentum components and disorder is weak, the results are essentially independent of~$\sigma$.

\begin{figure}
	\includegraphics{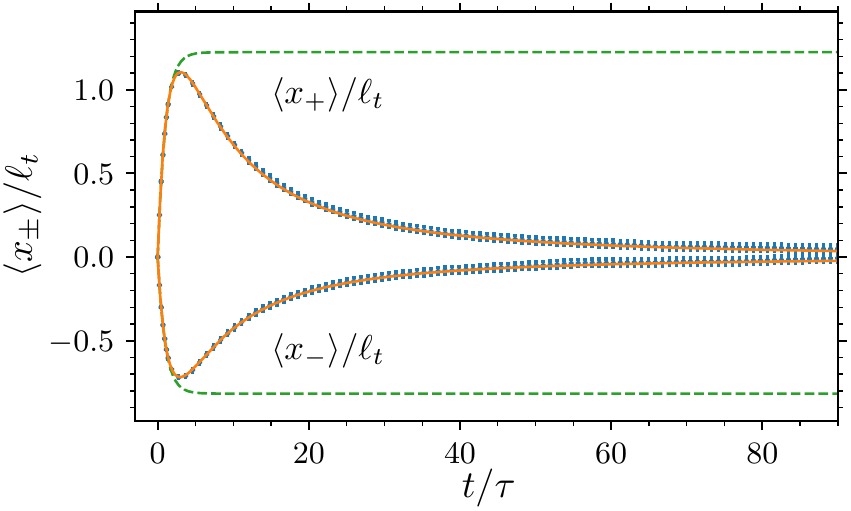}
	\caption{Temporal evolution of the center of mass~(\ref{eq:com_components}), for wavepackets launched with a positive (upper curve) and negative (lower curve) velocity. The blue symbols (with error bars representing statistical errors, with averaging over  40960 disorder configuration) are the quasi-exact numerical results. Even without time reversal invariance, we observe the full quantum boomerang effect -- the center of mass returns to the origin. The dashed lines present the classical solution~(\ref{eq:classical_solution}) which does not return to the origin. The orange solid lines are the prediction of our fully quantum theory \textit{\`a la} Berezinskii.	\label{fig:quantum_vs_boltzmann}
  }

\end{figure}
During the temporal evolution, only the two degenerate states are coupled by the weak disorder. The wavepacket can thus be written:
$\ket{\psi(t)} = \ket{\psi_+(t)}\otimes\ket{\chi_+} + \ket{\psi_-(t)}\otimes\ket{\chi_-}$, 
where $\psi_\pm(x,t)=\braket{x|\psi_\pm(t)}$ are the wave functions of the of the components propagating to the right (positive velocity, $+$ sign) and the left (negative veolcity, $-$ sign). This allows us to compute the CoM as:
\begin{equation}\label{eq:com_components}
    \langle x(t)\rangle = \int x\, \left(\overline{|\psi_+(x,t)|^2}+\overline{|\psi_-(x,t)|^2}\right)\ \diff{x}.
\end{equation}
{Due to non-orthogonality of the spinors $\ket{\chi}_\pm$, there exists also an interference term $\psi_+(x,t)\psi_-(x,t) + \text{c.c.}$. This term, however, is quickly oscillating, hence its contribution can be neglected in Eq.~(\ref{eq:com_components}).}

There are in total four possible elastic scattering events: $+\to +$, $+\to-$, $-\to-$ and $-\to+$.
The forward scattering events $+\to +$, $-\to -$ do not affect the dynamics of the CoM.  
Hence, in the further analysis, we use only the two scattering events where the direction of motion is changed. The associated scattering mean free times $\tau_+ =\tau_{+\to -}$ and $\tau_{-}=\tau_{-\to+}$ can be computed at weak disorder from the Fermi golden rule or, equivalently, from the Born approximation~\cite{Akkermans2007,Muller2011}:
\begin{equation}\label{eq:scattering_times}
    \frac{1}{\tau_+} = \frac{\eta\kappa}{\hbar^2 |v_-|}, \quad
    \frac{1}{\tau_-} = \frac{\eta\kappa}{\hbar^2 |v_+|}, 
\end{equation}
where $\kappa = |\braket{\chi_{+}|\chi_-}|^2$ is the spin-state overlap.

 \paragraph*{Classical solution--} 
The classical problem is governed by coupled Boltzmann equations 
$\partial_t f_\pm = - v_\pm \partial_x f_\pm \mp f_+/\tau_+ \pm f_-/\tau_-$, where $f_\pm(x,t)$ are the population densities. The subscript $\pm$ denotes the direction of propagation. These Boltzmann equations are easily solved~\cite{supple}.
We obtain:
\begin{equation}\label{eq:classical_solution}
 \langle x_\pm \rangle^\text{class.} = v_\pm \tau\left[1-\exp(-t/\tau)\right],   
\end{equation}
with $\tau = \tau_+\tau_-/(\tau_+ + \tau_-),$ a result very similar to the one obtained in the TRI case~\cite{Prat2019}. The particles, on average, travel a distance $|v_\pm| \tau$, then stop their evolution.

\paragraph*{Quantum numerics--}
We take a system with $\gamma=0.4$ and $\delta=\Omega=0.4$. The solution is obtained in a box large enough for the wavepacket not to touch the edges, $L=10000$, with a sufficiently small discretization $\Delta x =0.2$. The initial state's energy is chosen $E_0=0$, so that $k_-=-0.6453$ and $k_+=1.1850$. This also means that $\kappa=0.505$ and the velocities are $v_-=-0.5340$ and $v_+=0.8014$. We have used disorder strength $\eta=0.0049$, so that $\tau_-=323.61$, $\tau_+=215.66$, $\tau=129.42$, and the transport mean free path~\cite{supple} is $\ell_t=\tau\sqrt{|v_-| v_+}=84.67$.
For the time propagation, we have used the Chebyshev kernel method \cite{Tal-Ezer1984, Leforestier1991, Roche1997, Weisse2006}.

The results of the simulations are presented in Fig.~\ref{fig:quantum_vs_boltzmann}. Even though TRI is broken, we observe a \emph{perfect} QBE. Similarly to the TRI case, the CoM returns to the origin after the initial ballistic motion. The figure also includes classical solutions (dashed lines) obtained using Boltzmann equations. For a very short time, the classical solutions agree with quantum simulations. {Then Born approximation is sufficient to describe the quantum dynamics.} However, from $t\approx 3\tau,$ classical and quantum outcomes split, and the quantum particle returns to the origin.

The data displayed in Fig.~\ref{fig:quantum_vs_boltzmann} constitute a major result of this paper -- presence of the full QBE without TRI. All previous studies of the boomerang effect \cite{Prat2019, Janarek2020a, Tessieri2021, Sajjad2021} insisted on the importance of TRI (in addition to Anderson localization) for the boomerang effect. Here, we can give a negative answer to the question whether TRI is a necessary condition for the QBE existence. 

\paragraph*{Diagramatic approach--}
In addition to a purely numerical solution, Fig.~\ref{fig:quantum_vs_boltzmann}
displays theoretical predictions, described below, which perfectly agree
with numerics. We use an approach very similar to the one used for the TRI case in~\cite{Prat2019}: the Berezinskii diagrammatic technique~\cite{Berezinskii1996}.  The main difference with respect to earlier works~\cite{Berezinskii1996, Gogolin1976, Nakhmedov1987,  Wellens2016} is that there are more different scattering vertices due to the lack of TRI in our case.
The results are found as a Taylor series in powers of $t$ for the CoM positions~\cite{supple,Berezinskii_method:to_be_published}.

Here, for simplicity, we present solutions only for the positive initial velocity state. We obtain:
\begin{equation}\label{eq:short_time}
    \begin{split}
        &\langle x_+(t)\rangle = v_+ \tau\left[\frac{t}{\tau} - \frac{t^2}{2\tau^2} +  \frac{t^3}{6\tau^3} \right.  \\
        &\left. - \frac{(1+\Delta(4+\Delta(8+\Delta(4+\Delta))))t^4}{24(1+\Delta)^4 \tau^4}\right] + \mathcal{O}(t^5)
    \end{split}
\end{equation}
with $\Delta = |v_-/v_+|.$

This quantum solution agrees with the classical one up to third order:
\begin{equation}
\begin{split}
    \langle x_\pm(t) \rangle^\text{class.} = v_\pm \tau\left[\frac{t}{\tau} - \frac{t^2}{2\tau} + \frac{t^3}{6\tau^3} - \frac{t^4}{24\tau^4} \right] + \mathcal{O}(t^5),
\end{split}
\end{equation}
similarly to the TRI boomerang~\cite{Prat2019}. Likewise, we find a finite radius of convergence for the series in Eq.~(\ref{eq:short_time}), which seems to slightly depend on the value of $\Delta$.

A significant difference between our solution and the TRI case is that $\langle x_\pm(t)\rangle$ is no longer universal. Starting from the fourth order, all terms explicitly depend on $\Delta$. Of course, the TRI solution is fully recovered when $\Delta\!=\!1$.

To describe the center of mass evolution for intermediate times, we use a Pad\'e approximant~\cite{Baker1975} of the Taylor series.
The long-time scaling should be similar to the TRI system, i.e. $\langle x (t\!\gg\!\tau)\rangle \propto t^{-2}$, as also supported by the numerical evidence. While, in the TRI case, there is a more accurate asymptotic expression~\cite{Prat2019}:
$\langle x (t\!\gg\!\tau)\rangle\sim 64\ln(t/\tau)(\tau/t)^2,$ the derivation is much more difficult when TRI is broken~\cite{Berezinskii_method:to_be_published}, so we only know for sure the leading $t^{-2}$ behavior.  
Thus, we compute the CoM at any time as:
\begin{equation}
    \langle x_\pm (t)\rangle = \tau v_\pm \left(\frac{\tau}{t}\right)^2 \lim_{n\to\infty}R_n(t),
\end{equation}
where $R_n(t)$ is a diagonal Pad\'e approximant~\cite{Baker1975}, whose coefficients are calculated from the short-time Taylor series, Eq.~(\ref{eq:short_time}). To obtain high accuracy of the approximation, we use $n=30.$ There is however no visible difference with results obtained for lower $n$, e.g. $n=20$.

We compare the theoretical results with the numerical data in Fig.~\ref{fig:quantum_vs_boltzmann}.
The agreement is outstanding. We have performed a slight adjustment of the scattering rates $1/\tau_\pm$, given in the weak disorder limit by Eq.~(\ref{eq:scattering_times}) \cite{supple}. Indeed, higher order terms in the disorder strength $\eta$ are known to exist~\cite{Akkermans2007}, but they are small (of the order of 1\%)
and do not affect the structure of the Berezinskii method. Even without the adjustment, using the fully analytic expression (\ref{eq:scattering_times}), the agreement between the Berezinksii theory and the numerical results is excellent~\cite{supple}. This shows that the boomerang effect not only survives the breaking of TRI, but also that the time evolution of the CoM can be computed theoretically.

 \begin{figure}
 	\includegraphics{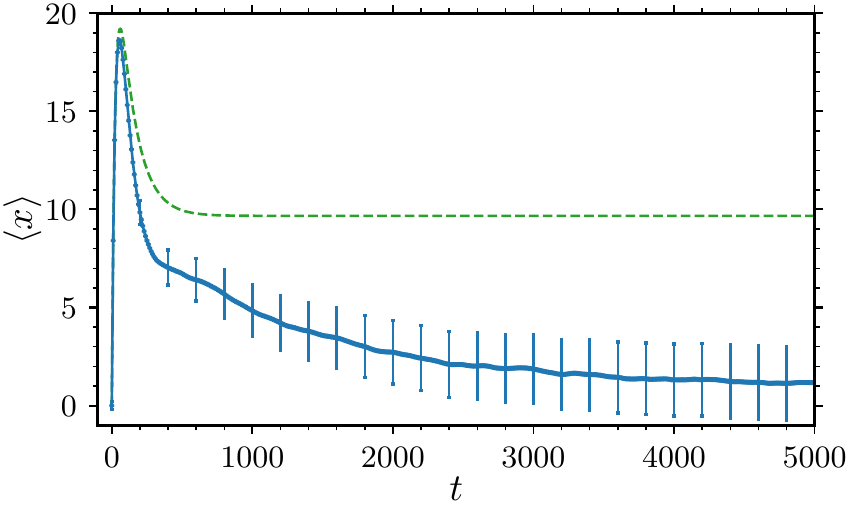}
 	\caption{Quantum boomerang effect in the 4-state scenario of Fig.~\ref{fig:spectrum_and_rbar}: $\gamma=0.4$ and $\delta=\Omega=0.4$, $E=0.75$, and $\eta=0.0225$ for a Gaussian state with initial momentum $k=0.68$ and width $\sigma=50$. The numerical calculation in blue shows that there is a full quantum boomerang effect at long time $\langle x(t\to \infty)\rangle = 0.$ In contrast, the classical solution of the coupled Boltzmann equations indicate a finite displacement of the center of mass of the wavepacket.
 	\label{fig:x_four_state}
 }
 \end{figure}

The QBE also exists in the 4-state scenario, when $H_0$ has 4 eigenstates  with the same energy $E_0,$ see Fig.~\ref{fig:spectrum_and_rbar}. As shown in Fig.~\ref{fig:x_four_state}, after an initial fast departure from the origin, again the CoM returns to the origin exhibiting a perfect QBE. The dashed line is the classical solution of the coupled Boltzmann equations~\cite{supple} which, amusingly, shows that the classical solution also reveals a tendency to return at short times, only then stopping at a non-zero distance from the origin. For this case, the diagrammatic approach, while feasible, would be very painful and is left for a possible future study.

Finally, we give an explanation of the observed QBE based on symmetry arguments. The product of the commuting parity operator $P$ and spinless time-reversal operator $T$ is such that $x\to -x,t\to -t,p\to p$ and does not touch the spin degree of freedom. It is an anti-unitary operator squaring to +1; it is thus \textit{not} a generalized time-reversal operator for the spin-1/2 system.
It modifies the disordered Hamiltonian so that:
\begin{equation}
	PT\ H\ (PT)^{-1} = H_0 + V(-x) = \tilde{H} \neq H.
\end{equation}
In an Anderson localized system, the infinite-time CoM position after evolving with $H$ may be obtained using the diagonal approximation:
\begin{equation}\label{eq:infinite_time_cmp}
	\begin{split}
		x^{H}(t=\infty) = \sum_i \bra{\phi_i}x\ket{\phi_i}|\braket{\phi_i | \psi_0}|^2=\\
		\sum_i \bra{\phi_i}x\ket{\phi_i}|\bra{\phi_i} TP\,PT \ket{\psi_0}|^2
	\end{split}
\end{equation}
where $\{\ket{\phi_i}\}$ is the eigenbasis of $H$ and  $\ket{\psi_0}$ is the initial state. The eigenstates of $\tilde{H}$ are  $\ket{\tilde{\phi}_i}= PT\ket{\phi_i}$. We have  $\bra{\phi_i}x\ket{\phi_i} = -\bra{\tilde{\phi}_i}x\ket{\tilde{\phi}_i}$. Inserting it in {Eq.~}(\ref{eq:infinite_time_cmp}), we get:
\begin{equation}
	 x^{H}(t=\infty) = -\sum_i \bra{\tilde{\phi}_i}x\ket{\tilde{\phi}_i} |\bra{\tilde{\phi}_i} PT\ket{\psi_0}|^2.
\end{equation}
Hence, if $PT\ket{\psi_0}=\ket{\psi_0}$ (which is the case for our initial Gaussian wavepackets), then $x^{H}(t=\infty) = -  x^{\tilde{H}}(t=\infty)$. For a single disorder realization, $ x^{H}(t=\infty)$ is generically nonzero. However, because $H$ and $\tilde{H}$ belong to the same statistical distribution of disorder with the same weight, after disorder averaging{, we obtain}
\begin{equation}
	\langle x(t=\infty)\rangle = \overline{ x^H(t=\infty) }= - \overline{ x^{\tilde{H}}(t=\infty) } =0,
\end{equation}
implying the full QBE. 

The Hamiltonian $H$ has no generalized TRI and is in the GUE symmetry class but still displays a full QBE, as exemplified in the inset of Fig.~\ref{fig:spectrum_and_rbar}.

\paragraph*{Summary.}We have demonstrated  the presence of the QBE in an exemplary spin-orbit system where all anti-unitary symmetries -- especially time reversal invariance -- are broken. We can give a sufficient condition for the appearance of the QBE in a general setting. The full QBE, i.e. $\langle x(t=\infty)\rangle = 0$, is present if there exists an unitary or antiunitary transformation $U$ satisfying the following conditions: {(i) the position operator is odd under the action of $U$: $UxU^{-1}\!=\!-x$, (ii) the disorder-free part of the Hamiltonian is symmetric under this transformation, $U H_0U^{-1} = H_0$, (iii) the transformed disorder $\tilde{V}=UVU^{-1}$ is a valid disorder realization with the same weight; (iv) and the initial state is symmetric under the transformation, $U\ket{\psi_0} = \ket{\psi_0}$}. For our Hamiltonian (\ref{eq:Hamiltonian_0}), such transformation is the product of the parity and spinless time reversal operators $U=PT$. 

By generalizing the Berezinskii diagrammatic approach to a system without TRI, we have computed the time evolution of the CoM in a quasi-analytic way. The theoretical result is in perfect agreement with the numerical data for the spin-orbit coupled model.
 Although in our work we have studied a one-dimensional system, similarly to the TRI counterpart of the phenomenon, we believe that the QBE is also present in higher-dimensional unitary systems.

 {After the original submission of this work, the preprint \cite{Macri22} appeared which had a partial overlap with our finding and generalizes QBE to certain non-Hermitian systems.}

\acknowledgments
We thank 
Tony Prat for discussions on the Berezinskii approach.  This research has been supported by the National Science Centre (Poland) under Project No. 2016/21/B/ST2/01086 (J.J.) and  2019/35/B/ST2/00034 (J.Z.). We acknowledge support of the Foundation for Polish Science (FNP)  through the first Polish-French Maria
Sklodowska-Pierre Curie reward received by D. Delande and
J. Zakrzewski. J.~Janarek  acknowledges also the support of French Embassy in Poland through the \emph{Bourse du Gouvernement Fran\c cais} program. We acknowledge the support of the PL-Grid structure which made numerical calculations possible.

%


\section{Supplementary Material for Quantum boomerang effect in systems without time reversal symmetry}

\renewcommand{\theequation}{S\arabic{equation}}
\renewcommand{\thefigure}{S\arabic{figure}}
\renewcommand{\bibnumfmt}[1]{[S#1]}

\subsection{Summary}
We give below a detailed treatment of the coupled Boltzmann equations governing the classical dynamics, the essential steps for the calculation of the center of mass of the quantum wavepacket, and a component by component comparison of the classical and quantum results.

\subsection{Classical calculation of the center of mass}
In the two-state scenario, the Boltzmann equations for the population densities $f_\pm(x,t)$ (where the $\pm$ index refers to atoms propagating with positive or negative velocity) are:
\begin{equation}\label{spin:eq:boltzmann_densities}
	\begin{dcases}
		\pdfrac{f_+}{t} &= -v_+\pdfrac{f_+}{x} - \frac{f_+}{\tau_+} + \frac{f_-}{\tau_-},\\[5pt]
		\pdfrac{f_-}{t} &= -v_-\pdfrac{f_-}{x} + \frac{f_+}{\tau_+} - \frac{f_-}{\tau_-}.
	\end{dcases}
\end{equation}
By integrating over the position $x,$ one obtains the populations themselves $n_\pm(t) = \int{f_\pm(x,t)\ \mathrm{d}x}$ which obey the coupled equations:
\begin{equation}\label{spin:eq:boltzmann_populations}
	\begin{dcases}
		\dfrac{n_+(t)}{t} &= \frac{n_-(t)}{\tau_-} - \frac{n_+(t)}{\tau_+},\\[5pt]
		\dfrac{n_-(t)}{t} &= \frac{n_+(t)}{\tau_+} - \frac{n_-(t)}{\tau_-}.
	\end{dcases}
\end{equation}
In the following, we consider the case of particles launched with a positive velocity. The case of
initial negative velocity can of course to treated in a similar manner.
The solution of Eq.~(\ref{spin:eq:boltzmann_populations}) is given by:
\begin{equation}\label{spin:eq:boltzmann_populations_solution}
	\begin{dcases}
		n_+(t) &=  \frac{\tau}{\tau_-} + \frac{\tau}{\tau_+} e^{-t/\tau}\\[5pt]
		n_-(t) &= \frac{\tau}{\tau_+}\bigl(1 - e^{-t/\tau}\bigr)
	\end{dcases}
\end{equation}
with:
\begin{equation}
	\tau = \frac{\tau_+\tau_-}{\tau_++\tau_-}
\end{equation}
The average velocity is $\langle v(t) \rangle = v_+n_+(t)+v_-n_-(t) = v_+ e^{-t/\tau},$ from which  
one obtains $\langle x(t)\rangle$, Eq.~(4) of the main text. 

For the comparison with the quantum results, it is useful to split the center of mass $\langle x(t) \rangle$ into the right and left propagating components $\langle x(t)\rangle_\pm,$ which can be computed from the population densities:
\begin{equation}
	\langle x(t) \rangle_\pm = \int{f_\pm(x,t)\ x\ \mathrm{d}x}
\end{equation} 
A simple integration by part gives the following set of equations:
\begin{equation}
	\begin{dcases}
		\dfrac{\langle x(t)\rangle_+}{t} &= v_+ n_+(t) - \frac{\langle x(t)\rangle_+}{\tau_+} + \frac{\langle x(t)\rangle_-}{\tau_-},\\[5pt]
		\dfrac{\langle x(t)\rangle_-}{t} &= v_- n_-(t) + \frac{\langle x(t)\rangle_+}{\tau_+} - \frac{\langle x(t)\rangle_-}{\tau_-}.
	\end{dcases}
\end{equation}
whose solution is:
\begin{equation}\label{eq:classical_two_state}
\begin{split}
    &\langle x(t)\rangle_+ = \\
    &\tau v_+\left (\frac{-2v_-}{v_+-v_-}\left(1-e^{t/\tau}\right) + \frac{v_+ + v_-}{v_+ - v_-}\left(\frac{t}{\tau}\right)e^{-t/\tau}\right), \\[10pt]
    &\langle x(t)\rangle_- = \\
    & \tau v_+\left(\frac{v_+ + v_-}{v_+ - v_-}\right)\left(1 -e^{-t/\tau} -\left(\frac{t}{\tau}\right)e^{-t/\tau}\right).
\end{split}
\end{equation}
Note that the sum  $\langle x(t)\rangle_+ + \langle x(t)\rangle_-$ is of course given by Eq.~(4) of the Letter.

Using the same set of equations, one can also compute the expectation values of $\langle x^2(t) \rangle.$ It involves the integrals $\int{f_\pm(x,t)\ x^2\ \mathrm{d}x}$ which can again be calculated with a simple integration by part. Not surprisingly, one obtains a diffusive behavior at long time $\mathrm{d}\langle x^2(t) \rangle/\mathrm{d}t = 2D$ with a diffusion coefficient given by $D=\tau \sqrt{|v_+v_-|}.$ This makes it possible to define a transport mean free path $\ell_t$ such that $D\tau=\ell_t^2$ and:
\begin{equation}
	\ell_t = \tau \sqrt{|v_+v_-|}
\end{equation} 

The four-state scenario can be investigated along the same lines, by splitting the density in four components associated with the four possible velocities. The calculation are straightforward but boring.
For simplicity, one can compute only the temporal evolution of the populations. They are governed by four coupled Boltzmann equations -- replacing the two coupled equations~(\ref{spin:eq:boltzmann_populations}) of the two-state scenario -- conveniently written using the scattering matrix $S$: $\mathrm{d}\bs{n}/\mathrm{d}t = S\,\bs{n}(t)$, where $\bs{n}(t) = (n_1(t), n_2(t), n_3(t), n_4(t))$. Among the four eigenvalues of $S$, one is zero and corresponds to the conservation of the total population. The three remaining eigenvalues are negative and control the relaxation towards the long-time steady state. The total center of mass is computed as $\langle x(t)\rangle = \int_0^t \bs{n}(t')\cdot\bs{v}\, \diff{t'}$, where $\bs{v}=(v_1, v_2, v_3, v_4)$ is the vector of population velocities.

\subsection{Quantum calculation of the center of mass}
\begin{figure}
    \centering
    \includegraphics{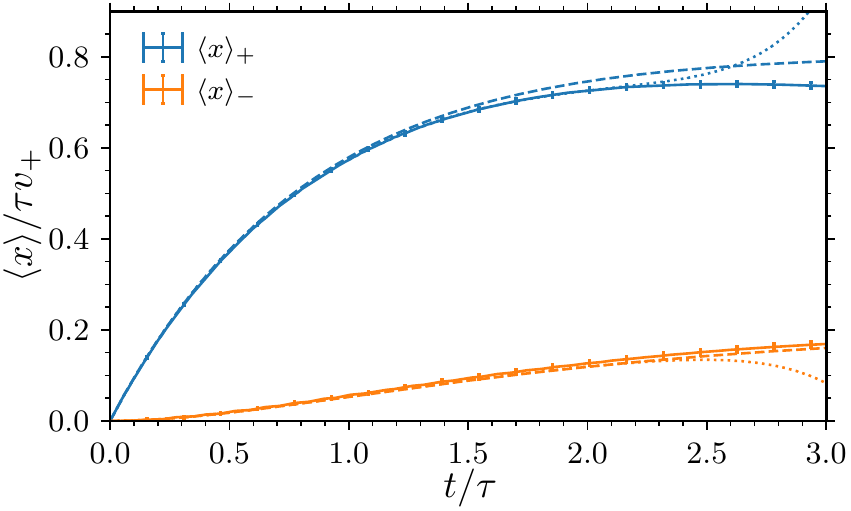}
    \caption{Comparison of the numerical results (solid lines with error bars), classical solutions (\ref{eq:classical_two_state}) (dashed lines), and the quantum expansion calculated up to the $(t/\tau)^{11}$ order for $\Delta = 2/3$ (dotted lines) for $\langle x(t)\rangle_+$ and $\langle x(t)\rangle_-$. At short time, the classical and quantum solutions agree very well with the numerical data, the quantum expansion showing better agreement with the numerical data for $t < 2.3\tau$.
    \label{fig:short_time_solution}}
\end{figure}

To compute the center of mass of the quantum wavepacket, we express it using the retarded and advanced Green's functions, starting from the general formula for the center of mass:
\begin{equation}\label{sup:eq:com_0}
    \langle x(t)\rangle = \int\diff{x} x\,\overline{|\psi(x,t)|^2}.
\end{equation}
Let $G^{R/A}(x,x',E)$ denote the retarded/advanced Green's functions at energy $E.$ We can conveniently express the center of mass, Eq.~(\ref{sup:eq:com_0}), in the frequency space~\cite{Akkermans2007}:
\begin{equation}\label{sup:eq:com_1}
\begin{split}
    \langle x(\omega)\rangle &= \frac{1}{2\pi} \int\diff{x}\diff{x'}\diff{x''}\diff{E} x \times\\ &\overline{G^R(x,x',E)G^A(x'',x,E-\omega)} \Psi_0(x')\Psi^*_0(x''),
\end{split}
\end{equation}
where $\Psi_0(x)$ is the initial state of the system. The central object of interest in Eq.~(\ref{sup:eq:com_1}) is the average product of the Green's functions, which describes the transport properties of the system. Equation~(\ref{sup:eq:com_1}) is general and does not depend on the details of the system, e.g., the TRI breaking or spin.

To compute the product of Green's functions, we start from the Born series~\cite{Akkermans2007},
\begin{equation}
\begin{split}
    &G^{R/A}(x,x',E) = G_0^{R/A}(x,x',E)\\ &+ \int\diff{x_1} G_0^{R/A}(x,x_1,E)V(x_1)G_0^{R/A}(x_1,x',E)+\ldots
\end{split}
\end{equation}
so that $G^R(x,x',\epsilon)G^A(x'',x,E-\omega)$ involves only products of free Green's functions $G_0^{R/A}(x,x',E)$ and disordered potentials, that is multiply scattered paths.

The average product of the Green's functions includes all possible multiply scattered paths. In the Berezinskii diagrammatic technique, each path is represented as a diagram, and specific rules exist to classify all possible diagrams~\cite{Berezinskii1996,Prat2019,Prat2017a}. Within this approach, we can sum all relevant contributions to the average product of the Green's functions. This approach goes beyond the Born approximation and beyond the standard diagrammatic approach used for weakly disordered systems~\cite{Akkermans2007,Muller2011}, as it incorporates diagrams which are neither "ladder" diagrams, nor "maximally crossed" ones. In our model, we assume that the disordered potential is a Gaussian uncorrelated disorder. Thanks to the Wick theorem, only diagrams with paired scattering events contribute, greatly simplifying calculations. Note however that this pairing does not necessarily involves one retarded and one advanced Green functions -- as is the case for the standard diagrammatic approach~\cite{Akkermans2007,Muller2011} -- but may involve two Green functions of the same kind.

The complete calculation involves many details which will be presented in a separate forthcoming publication~\cite{Berezinskii_method:to_be_published}. We here briefly describe the important steps.
 
To sum all diagrams, we analyze their structure. The most important ingredient are so-called scattering vertices, which describe all possible scattering events in the system. The scattering vertices are system dependent, i.e., they depend on the details of the system like the lack of TRI. The final analysis of the diagrams leads us to:
\begin{equation}\label{sup:eq:com_quantum}
\begin{split}
    \langle x(t)\rangle_+ =& \tau_+ \int \frac{\diff{\omega}}{2\pi} e^{-i\omega t} \sum_m \left(L_m S^0_m + L_{m+1} S^1_m\right),\\
    \langle x(t)\rangle_- =& -\tau_- \int \frac{\diff{\omega}}{2\pi} e^{-i\omega t} \sum_m \left(L_mS^2_m + L_m S^3_m\right),
\end{split}
\end{equation}
where $S^i_m$ and $Q^i_m$ obey the system of equations~(\ref{sup:eq:full_set_s_q}) with $\nu = \omega(v_+-v_-)/\kappa \eta$, $s\!=\!2-\frac{2}{\kappa}\!+\!i\nu$, $\beta_m = (2\kappa m^2+2m+1)/|v_- v_+|$, $\ell^* = |v_+ v_-|/(\kappa \eta)$, and  $L_m(s) = -s\Gamma(m+1)\Psi(m+1,2;-s)$, where $\Psi(a,b;z)$ is the confluent hypergeometric function of the second kind~\cite{Abramowitz1966}.

\begin{figure}
    \centering
    \includegraphics{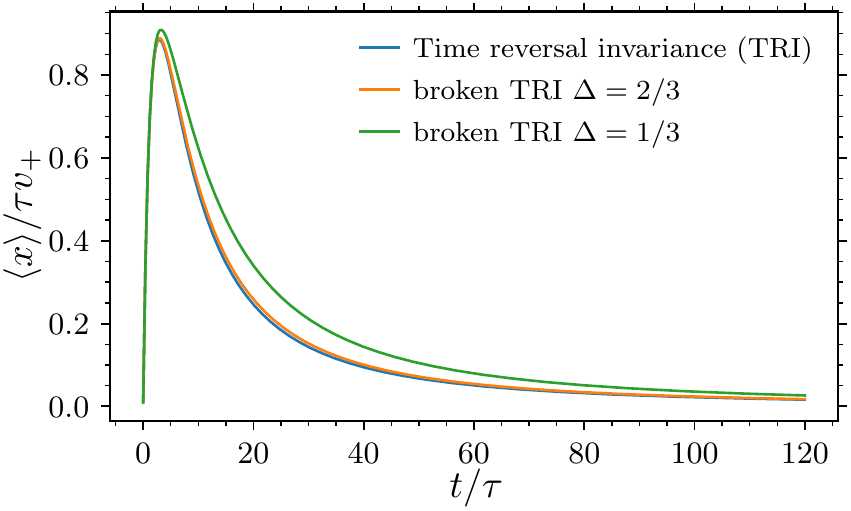}
    \caption{Comparison of the Pad\'e approximated quantum results for the TRI system (blue line, $\Delta=1$) and broken TRI systems with different values of velocities ratio $\Delta=|v_-/v_+|$ ($\Delta =2/3$ orange line, $\Delta=1/3$ green line). For a moderate TRI breaking such as $\Delta=2/3,$ the difference with the TRI case $\Delta=1$ is almost invisible. For smaller $\Delta,$ some deviation is observed, but the overall shape is very similar. 
    \label{fig:new_pade}}
\end{figure}
At short times, $S^i_m(\omega)$ and  $Q^i_m(\omega)$ can be found with the help of expansions $S^i_m(\omega)=\sum_n s^i_{m,n}/(i\omega)^n$, and $Q^i_m(\omega)=\sum_n q^i_{m,n}/(i\omega)^n$.  The series can be computed to arbitrary order in $1/\omega$, enabling calculation of $\langle x(t)\rangle$ as a Taylor expansion in powers of $t.$  In Fig.~\ref{fig:short_time_solution}, we present a comparison of the numerical results with classical and quantum (calculated up to the $(t/\tau)^{11}$ order) theoretical solutions. For very short times, the data and both theoretical approaches agree very well. At longer time, around $t=1.5\tau,$ the classical prediction starts to differ from both the quantum one and the numerical data. Due to the finite radius of convergence of the series, the quantum expansion deviates from the data for longer times, $t>2.5\tau$.

\begin{widetext}
\begin{equation}\label{sup:eq:full_set_s_q}
    \begin{gathered}
        \ell^* Q^0_m+i\nu\left(m-\frac{v_-}{v_+-v_-}\right)S^0_m- \eta \ell^*  \beta_m S^0_m + m^2S^0_{m-1} +(m+1)^2S^0_{m+1} = 0, \\
        L_m+i\nu \left(m-\frac{v_-}{v_+-v_-}\right) Q^0_m-\eta\ell^*\beta_mQ^0_m+m^2Q^0_{m-1}+(m+1)^2Q^0_{m+1} =0, \\[10pt]
        -\ell^*Q^1_m+i\nu\left(m + \frac{v_+}{v_+-v_-}\right)S^1_m - \eta\ell^*\beta_m S^1_m + m^2S^1_{m-1} + (m+1)^2 S^1_{m+1} = 0,\\
        L_{m+1} + i\nu \left(m + \frac{v_+}{v_+-v_-}\right)Q^1_m-\eta\ell^*\beta_mQ^1_m+m^2Q^1_{m-1} + (m+1)^2Q^1_{m+1} = 0,\\[10pt]
        \ell^* Q^2_m+i\nu\left(m-\frac{v_-}{v_+-v_-}\right) S^2_m - \eta\ell^*\beta_mS^2_m +m^2S^2_{m-1} + (m+1)^2S^2_{m+1} = 0,\\
        L_{m+1} + i\nu \left(m-\frac{v_-}{v_+-v_-}\right)Q^2_m - \eta\ell^*\beta_mQ^2_m + m^2Q^2_{m-1} + (m+1)^2Q^2_{m+1} =0,\\[10pt]
        -\ell^*Q^3_m+i\nu\left(1+\frac{v_+}{v_+-v_-}\right)S^3_m-\eta\ell^*\beta_mS^3_m + m^2S^3_{m-1} + (m+1)^2S^3_{m+1} = 0,\\
        R_{m+1}+i\nu\left(1+\frac{v_+}{v_+-v_-}\right)Q^3_m-\eta\ell^*\beta_mQ^3_m+m^2Q^3_{m-1} + (m+1)^2Q^3_{m+1} = 0.
    \end{gathered}
\end{equation}
\end{widetext}

\subsection{Quantum solutions with Pad\'e approximant}

\begin{figure}
    \centering
    \includegraphics{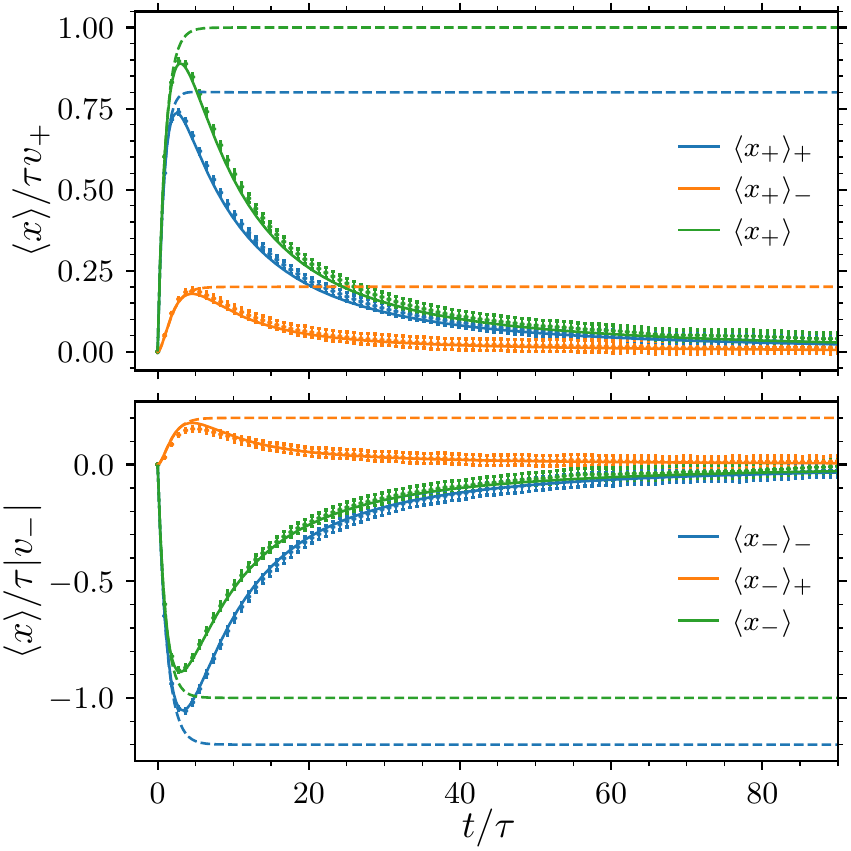}
    \caption{Comparison of the numerical data (points with error bars, for details see text) with the quantum solutions, Eq.~(\ref{sup:eq:com_quantum}) (solid lines) and the classical solutions, Eq.~(\ref{eq:classical_two_state}) (dashed lines, computed for initial positive and negative velocities). In contrast with Fig.~2 of the Letter, the theoretical quantum prediction is presented without any adjustment of the time and length scales, using the analytical Born expressions. Even without any adjustment, which takes into account the differences between the Born mean free times/paths and true ones, the agreement is excellent.
    \label{fig:cmp_vs_pade_without_fit}}
\end{figure}

From the short time expansion, we can compute the center of mass at any time by using a Pad\'e approximant, as explained in the main text of the Letter. The same technique is used to compute the forward and backward scattered components of the center of mass. It turns out that, in the studied case with $\Delta=|v_-/v_+| = 2/3$, the total center of mass $\langle x_\pm(t)\rangle$ does not differ significantly from the TRI boomerang curve. For stronger TRI violation (smaller $\Delta$), the difference becomes larger, see Fig.~\ref{fig:new_pade}. This, once again, shows that contrary to the TRI case, the center of mass time dependence $\langle x_\pm(t)\rangle$ is no longer universal.

From previous works on the QBE \cite{Prat2019, Tessieri2021, Janarek2020a}, it is known that the boomerang curves need a very high number of disorder realizations to have small error bars. Moreover, the theoretical calculation is valid only in the weak disorder limit. The net difference between the theoretical and numerical results appears as a difference between the exact mean free time and length (present in the simulations) and their values computed with the Born approximation (used in theoretical calculations). The difference can be captured using a simple adjustment of the time and length scales. If we split the two components $\langle x_+\rangle_+$ and $\langle x_+\rangle_-$ -- respectively corresponding to the contributions of the $\psi_+(x,t)$ and $\psi_-(x,t)$ of the wavepacket -- we find for an initial positive velocity, $k_+(\tau v_+)\approx 120$:
\begin{equation}
  \begin{aligned}
    &\langle x_+\rangle:\quad &\tau_\text{fit}=1.03\tau,\quad &\ell_\text{fit} = 1.015 \tau v_+ \\
    &\langle x_+\rangle_+:\quad &\tau_\text{fit}=1.04\tau,\quad &\ell_\text{fit} = 1.01 \tau v_+ \\
    &\langle x_+\rangle_-:\quad &\tau_\text{fit}=1.03\tau,\quad &\ell_\text{fit} = 1.07 \tau v_+\\
  \end{aligned}
\end{equation}
For the negative initial velocity, $k_-(\tau |v_-|)\approx 45$:
\begin{equation}
  \begin{aligned}
    &\langle x_-\rangle:\quad &\tau_\text{fit}=0.99\tau,\quad &\ell_\text{fit} = 0.99 \tau |v_-| \\
    &\langle x_-\rangle_+:\quad &\tau_\text{fit}=1.13\tau,\quad &\ell_\text{fit} = 0.87 \tau |v_-| \\
    &\langle x_-\rangle_-:\quad &\tau_\text{fit}=1.01\tau,\quad &\ell_\text{fit} = 1.00 \tau |v_-|
  \end{aligned}
\end{equation}

The differences are very small, on the order of a few \%. The largest ones are observed for the backscattered parts, i.e. $\langle x_+\rangle_-$ and $\langle x_-\rangle_+$. This is because the backscattering processes are intrinsically of higher order, and higher differences between exact and Born parameters are expected. Figure~\ref{fig:cmp_vs_pade_without_fit} presents a full comparison of all contributions to the total center of mass for both initial states. It is clear that even without any adjustment of the time and length scales, that is using all quantities analytically computed with the Born approximation, the agreement between the numerical results and the quantum theoretical solutions is very good.


\begin{thebibliography}{36}%
\makeatletter
\providecommand \@ifxundefined [1]{%
 \@ifx{#1\undefined}
}%
\providecommand \@ifnum [1]{%
 \ifnum #1\expandafter \@firstoftwo
 \else \expandafter \@secondoftwo
 \fi
}%
\providecommand \@ifx [1]{%
 \ifx #1\expandafter \@firstoftwo
 \else \expandafter \@secondoftwo
 \fi
}%
\providecommand \natexlab [1]{#1}%
\providecommand \enquote  [1]{``#1''}%
\providecommand \bibnamefont  [1]{#1}%
\providecommand \bibfnamefont [1]{#1}%
\providecommand \citenamefont [1]{#1}%
\providecommand \href@noop [0]{\@secondoftwo}%
\providecommand \href [0]{\begingroup \@sanitize@url \@href}%
\providecommand \@href[1]{\@@startlink{#1}\@@href}%
\providecommand \@@href[1]{\endgroup#1\@@endlink}%
\providecommand \@sanitize@url [0]{\catcode `\\12\catcode `\$12\catcode
  `\&12\catcode `\#12\catcode `\^12\catcode `\_12\catcode `\%12\relax}%
\providecommand \@@startlink[1]{}%
\providecommand \@@endlink[0]{}%
\providecommand \url  [0]{\begingroup\@sanitize@url \@url }%
\providecommand \@url [1]{\endgroup\@href {#1}{\urlprefix }}%
\providecommand \urlprefix  [0]{URL }%
\providecommand \Eprint [0]{\href }%
\providecommand \doibase [0]{https://doi.org/}%
\providecommand \selectlanguage [0]{\@gobble}%
\providecommand \bibinfo  [0]{\@secondoftwo}%
\providecommand \bibfield  [0]{\@secondoftwo}%
\providecommand \translation [1]{[#1]}%
\providecommand \BibitemOpen [0]{}%
\providecommand \bibitemStop [0]{}%
\providecommand \bibitemNoStop [0]{.\EOS\space}%
\providecommand \EOS [0]{\spacefactor3000\relax}%
\providecommand \BibitemShut  [1]{\csname bibitem#1\endcsname}%
\let\auto@bib@innerbib\@empty
\bibitem [{\citenamefont {Anderson}(1958)}]{Anderson1958}%
  \BibitemOpen
  \bibfield  {author} {\bibinfo {author} {\bibfnamefont {P.~W.}\ \bibnamefont
  {Anderson}},\ }\bibfield  {title} {\bibinfo {title} {{Absence of diffusion in
  certain random lattices}},\ }\href {https://doi.org/10.1103/PhysRev.109.1492}
  {\bibfield  {journal} {\bibinfo  {journal} {Phys. Rev.}\ }\textbf {\bibinfo
  {volume} {109}},\ \bibinfo {pages} {1492} (\bibinfo {year}
  {1958})}\BibitemShut {NoStop}%
\bibitem [{\citenamefont {Roati}\ \emph {et~al.}(2008)\citenamefont {Roati},
  \citenamefont {D'Errico}, \citenamefont {Fallani}, \citenamefont {Fattori},
  \citenamefont {Fort}, \citenamefont {Zaccanti}, \citenamefont {Modugno},
  \citenamefont {Modugno},\ and\ \citenamefont {Inguscio}}]{Roati2008}%
  \BibitemOpen
  \bibfield  {author} {\bibinfo {author} {\bibfnamefont {G.}~\bibnamefont
  {Roati}}, \bibinfo {author} {\bibfnamefont {C.}~\bibnamefont {D'Errico}},
  \bibinfo {author} {\bibfnamefont {L.}~\bibnamefont {Fallani}}, \bibinfo
  {author} {\bibfnamefont {M.}~\bibnamefont {Fattori}}, \bibinfo {author}
  {\bibfnamefont {C.}~\bibnamefont {Fort}}, \bibinfo {author} {\bibfnamefont
  {M.}~\bibnamefont {Zaccanti}}, \bibinfo {author} {\bibfnamefont
  {G.}~\bibnamefont {Modugno}}, \bibinfo {author} {\bibfnamefont
  {M.}~\bibnamefont {Modugno}},\ and\ \bibinfo {author} {\bibfnamefont
  {M.}~\bibnamefont {Inguscio}},\ }\bibfield  {title} {\bibinfo {title}
  {{Anderson localization of a non-interacting Bose-Einstein condensate}},\
  }\href {https://doi.org/10.1038/nature07071} {\bibfield  {journal} {\bibinfo
  {journal} {Nature}\ }\textbf {\bibinfo {volume} {453}},\ \bibinfo {pages}
  {895} (\bibinfo {year} {2008})}\BibitemShut {NoStop}%
\bibitem [{\citenamefont {Billy}\ \emph {et~al.}(2008)\citenamefont {Billy},
  \citenamefont {Josse}, \citenamefont {Zuo}, \citenamefont {Bernard},
  \citenamefont {Hambrecht}, \citenamefont {Lugan}, \citenamefont
  {Cl{\'{e}}ment}, \citenamefont {Sanchez-Palencia}, \citenamefont {Bouyer},\
  and\ \citenamefont {Aspect}}]{Billy2008}%
  \BibitemOpen
  \bibfield  {author} {\bibinfo {author} {\bibfnamefont {J.}~\bibnamefont
  {Billy}}, \bibinfo {author} {\bibfnamefont {V.}~\bibnamefont {Josse}},
  \bibinfo {author} {\bibfnamefont {Z.}~\bibnamefont {Zuo}}, \bibinfo {author}
  {\bibfnamefont {A.}~\bibnamefont {Bernard}}, \bibinfo {author} {\bibfnamefont
  {B.}~\bibnamefont {Hambrecht}}, \bibinfo {author} {\bibfnamefont
  {P.}~\bibnamefont {Lugan}}, \bibinfo {author} {\bibfnamefont
  {D.}~\bibnamefont {Cl{\'{e}}ment}}, \bibinfo {author} {\bibfnamefont
  {L.}~\bibnamefont {Sanchez-Palencia}}, \bibinfo {author} {\bibfnamefont
  {P.}~\bibnamefont {Bouyer}},\ and\ \bibinfo {author} {\bibfnamefont
  {A.}~\bibnamefont {Aspect}},\ }\bibfield  {title} {\bibinfo {title} {{Direct
  observation of Anderson localization of matter waves in a controlled
  disorder}},\ }\href {https://doi.org/10.1038/nature07000} {\bibfield
  {journal} {\bibinfo  {journal} {Nature}\ }\textbf {\bibinfo {volume} {453}},\
  \bibinfo {pages} {891} (\bibinfo {year} {2008})}\BibitemShut {NoStop}%
\bibitem [{\citenamefont {Jendrzejewski}\ \emph {et~al.}(2012)\citenamefont
  {Jendrzejewski}, \citenamefont {M{\"{u}}ller}, \citenamefont {Richard},
  \citenamefont {Date}, \citenamefont {Plisson}, \citenamefont {Bouyer},
  \citenamefont {Aspect},\ and\ \citenamefont {Josse}}]{Jendrzejewski2012}%
  \BibitemOpen
  \bibfield  {author} {\bibinfo {author} {\bibfnamefont {F.}~\bibnamefont
  {Jendrzejewski}}, \bibinfo {author} {\bibfnamefont {K.}~\bibnamefont
  {M{\"{u}}ller}}, \bibinfo {author} {\bibfnamefont {J.}~\bibnamefont
  {Richard}}, \bibinfo {author} {\bibfnamefont {A.}~\bibnamefont {Date}},
  \bibinfo {author} {\bibfnamefont {T.}~\bibnamefont {Plisson}}, \bibinfo
  {author} {\bibfnamefont {P.}~\bibnamefont {Bouyer}}, \bibinfo {author}
  {\bibfnamefont {A.}~\bibnamefont {Aspect}},\ and\ \bibinfo {author}
  {\bibfnamefont {V.}~\bibnamefont {Josse}},\ }\bibfield  {title} {\bibinfo
  {title} {{Coherent backscattering of ultracold atoms}},\ }\href
  {https://doi.org/10.1103/PhysRevLett.109.195302} {\bibfield  {journal}
  {\bibinfo  {journal} {Phys. Rev. Lett.}\ }\textbf {\bibinfo {volume} {109}},\
  \bibinfo {pages} {195302} (\bibinfo {year} {2012})}\BibitemShut {NoStop}%
\bibitem [{\citenamefont {Manai}\ \emph {et~al.}(2015)\citenamefont {Manai},
  \citenamefont {Cl{\'{e}}ment}, \citenamefont {Chicireanu}, \citenamefont
  {Hainaut}, \citenamefont {Garreau}, \citenamefont {Szriftgiser},\ and\
  \citenamefont {Delande}}]{Manai2015}%
  \BibitemOpen
  \bibfield  {author} {\bibinfo {author} {\bibfnamefont {I.}~\bibnamefont
  {Manai}}, \bibinfo {author} {\bibfnamefont {J.-F.}\ \bibnamefont
  {Cl{\'{e}}ment}}, \bibinfo {author} {\bibfnamefont {R.}~\bibnamefont
  {Chicireanu}}, \bibinfo {author} {\bibfnamefont {C.}~\bibnamefont {Hainaut}},
  \bibinfo {author} {\bibfnamefont {J.~C.}\ \bibnamefont {Garreau}}, \bibinfo
  {author} {\bibfnamefont {P.}~\bibnamefont {Szriftgiser}},\ and\ \bibinfo
  {author} {\bibfnamefont {D.}~\bibnamefont {Delande}},\ }\bibfield  {title}
  {\bibinfo {title} {{Experimental Observation of Two-Dimensional Anderson
  Localization with the Atomic Kicked Rotor}},\ }\href
  {https://doi.org/10.1103/PhysRevLett.115.240603} {\bibfield  {journal}
  {\bibinfo  {journal} {Phys. Rev. Lett.}\ }\textbf {\bibinfo {volume} {115}},\
  \bibinfo {pages} {240603} (\bibinfo {year} {2015})},\ \Eprint
  {https://arxiv.org/abs/1504.04987} {arXiv:1504.04987} \BibitemShut {NoStop}%
\bibitem [{\citenamefont {Semeghini}\ \emph {et~al.}(2015)\citenamefont
  {Semeghini}, \citenamefont {Landini}, \citenamefont {Castilho}, \citenamefont
  {Roy}, \citenamefont {Spagnolli}, \citenamefont {Trenkwalder}, \citenamefont
  {Fattori}, \citenamefont {Inguscio},\ and\ \citenamefont
  {Modugno}}]{Semeghini2015}%
  \BibitemOpen
  \bibfield  {author} {\bibinfo {author} {\bibfnamefont {G.}~\bibnamefont
  {Semeghini}}, \bibinfo {author} {\bibfnamefont {M.}~\bibnamefont {Landini}},
  \bibinfo {author} {\bibfnamefont {P.}~\bibnamefont {Castilho}}, \bibinfo
  {author} {\bibfnamefont {S.}~\bibnamefont {Roy}}, \bibinfo {author}
  {\bibfnamefont {G.}~\bibnamefont {Spagnolli}}, \bibinfo {author}
  {\bibfnamefont {A.}~\bibnamefont {Trenkwalder}}, \bibinfo {author}
  {\bibfnamefont {M.}~\bibnamefont {Fattori}}, \bibinfo {author} {\bibfnamefont
  {M.}~\bibnamefont {Inguscio}},\ and\ \bibinfo {author} {\bibfnamefont
  {G.}~\bibnamefont {Modugno}},\ }\bibfield  {title} {\bibinfo {title}
  {{Measurement of the mobility edge for 3D Anderson localization}},\ }\href
  {https://doi.org/10.1038/nphys3339} {\bibfield  {journal} {\bibinfo
  {journal} {Nat. Phys.}\ }\textbf {\bibinfo {volume} {11}},\ \bibinfo {pages}
  {554} (\bibinfo {year} {2015})}\BibitemShut {NoStop}%
\bibitem [{\citenamefont {Hu}\ \emph {et~al.}(2008)\citenamefont {Hu},
  \citenamefont {Strybulevych}, \citenamefont {Page}, \citenamefont
  {Skipetrov},\ and\ \citenamefont {van Tiggelen}}]{Hu2008}%
  \BibitemOpen
  \bibfield  {author} {\bibinfo {author} {\bibfnamefont {H.}~\bibnamefont
  {Hu}}, \bibinfo {author} {\bibfnamefont {A.}~\bibnamefont {Strybulevych}},
  \bibinfo {author} {\bibfnamefont {J.~H.}\ \bibnamefont {Page}}, \bibinfo
  {author} {\bibfnamefont {S.~E.}\ \bibnamefont {Skipetrov}},\ and\ \bibinfo
  {author} {\bibfnamefont {B.~A.}\ \bibnamefont {van Tiggelen}},\ }\bibfield
  {title} {\bibinfo {title} {{Localization of ultrasound in a three-dimensional
  elastic network}},\ }\href {https://doi.org/10.1038/nphys1101} {\bibfield
  {journal} {\bibinfo  {journal} {Nat. Phys.}\ }\textbf {\bibinfo {volume}
  {4}},\ \bibinfo {pages} {945} (\bibinfo {year} {2008})},\ \Eprint
  {https://arxiv.org/abs/0805.1502} {arXiv:0805.1502} \BibitemShut {NoStop}%
\bibitem [{\citenamefont {Chabanov}\ \emph {et~al.}(2000)\citenamefont
  {Chabanov}, \citenamefont {Stoytchev},\ and\ \citenamefont
  {Genack}}]{Chabanov2000}%
  \BibitemOpen
  \bibfield  {author} {\bibinfo {author} {\bibfnamefont {A.~A.}\ \bibnamefont
  {Chabanov}}, \bibinfo {author} {\bibfnamefont {M.}~\bibnamefont
  {Stoytchev}},\ and\ \bibinfo {author} {\bibfnamefont {A.~Z.}\ \bibnamefont
  {Genack}},\ }\bibfield  {title} {\bibinfo {title} {{Statistical signatures of
  photon localization}},\ }\href {https://doi.org/10.1038/35009055} {\bibfield
  {journal} {\bibinfo  {journal} {Nature}\ }\textbf {\bibinfo {volume} {404}},\
  \bibinfo {pages} {850} (\bibinfo {year} {2000})}\BibitemShut {NoStop}%
\bibitem [{\citenamefont {Schwartz}\ \emph {et~al.}(2007)\citenamefont
  {Schwartz}, \citenamefont {Bartal}, \citenamefont {Fishman},\ and\
  \citenamefont {Segev}}]{Schwartz2007}%
  \BibitemOpen
  \bibfield  {author} {\bibinfo {author} {\bibfnamefont {T.}~\bibnamefont
  {Schwartz}}, \bibinfo {author} {\bibfnamefont {G.}~\bibnamefont {Bartal}},
  \bibinfo {author} {\bibfnamefont {S.}~\bibnamefont {Fishman}},\ and\ \bibinfo
  {author} {\bibfnamefont {M.}~\bibnamefont {Segev}},\ }\bibfield  {title}
  {\bibinfo {title} {{Transport and Anderson localization in disordered
  two-dimensional photonic lattices}},\ }\href
  {https://doi.org/10.1038/nature05623} {\bibfield  {journal} {\bibinfo
  {journal} {Nature}\ }\textbf {\bibinfo {volume} {446}},\ \bibinfo {pages}
  {52} (\bibinfo {year} {2007})}\BibitemShut {NoStop}%
\bibitem [{\citenamefont {Prat}\ \emph {et~al.}(2019)\citenamefont {Prat},
  \citenamefont {Delande},\ and\ \citenamefont {Cherroret}}]{Prat2019}%
  \BibitemOpen
  \bibfield  {author} {\bibinfo {author} {\bibfnamefont {T.}~\bibnamefont
  {Prat}}, \bibinfo {author} {\bibfnamefont {D.}~\bibnamefont {Delande}},\ and\
  \bibinfo {author} {\bibfnamefont {N.}~\bibnamefont {Cherroret}},\ }\bibfield
  {title} {\bibinfo {title} {{Quantum boomeranglike effect of wave packets in
  random media}},\ }\href {https://doi.org/10.1103/PhysRevA.99.023629}
  {\bibfield  {journal} {\bibinfo  {journal} {Phys. Rev. A}\ }\textbf {\bibinfo
  {volume} {99}},\ \bibinfo {pages} {023629} (\bibinfo {year}
  {2019})}\BibitemShut {NoStop}%
\bibitem [{\citenamefont {Tessieri}\ \emph {et~al.}(2021)\citenamefont
  {Tessieri}, \citenamefont {Akdeniz}, \citenamefont {Cherroret}, \citenamefont
  {Delande},\ and\ \citenamefont {Vignolo}}]{Tessieri2021}%
  \BibitemOpen
  \bibfield  {author} {\bibinfo {author} {\bibfnamefont {L.}~\bibnamefont
  {Tessieri}}, \bibinfo {author} {\bibfnamefont {Z.}~\bibnamefont {Akdeniz}},
  \bibinfo {author} {\bibfnamefont {N.}~\bibnamefont {Cherroret}}, \bibinfo
  {author} {\bibfnamefont {D.}~\bibnamefont {Delande}},\ and\ \bibinfo {author}
  {\bibfnamefont {P.}~\bibnamefont {Vignolo}},\ }\bibfield  {title} {\bibinfo
  {title} {{Quantum boomerang effect: Beyond the standard Anderson model}},\
  }\href {https://doi.org/10.1103/PhysRevA.103.063316} {\bibfield  {journal}
  {\bibinfo  {journal} {Phys. Rev. A}\ }\textbf {\bibinfo {volume} {103}},\
  \bibinfo {pages} {063316} (\bibinfo {year} {2021})}\BibitemShut {NoStop}%
\bibitem [{\citenamefont {Sajjad}\ \emph {et~al.}(2022)\citenamefont {Sajjad},
  \citenamefont {Tanlimco}, \citenamefont {Mas}, \citenamefont {Cao},
  \citenamefont {Nolasco-Martinez}, \citenamefont {Simmons}, \citenamefont
  {Santos}, \citenamefont {Vignolo}, \citenamefont {Macr\`{\i}},\ and\
  \citenamefont {Weld}}]{Sajjad2021}%
  \BibitemOpen
  \bibfield  {author} {\bibinfo {author} {\bibfnamefont {R.}~\bibnamefont
  {Sajjad}}, \bibinfo {author} {\bibfnamefont {J.~L.}\ \bibnamefont
  {Tanlimco}}, \bibinfo {author} {\bibfnamefont {H.}~\bibnamefont {Mas}},
  \bibinfo {author} {\bibfnamefont {A.}~\bibnamefont {Cao}}, \bibinfo {author}
  {\bibfnamefont {E.}~\bibnamefont {Nolasco-Martinez}}, \bibinfo {author}
  {\bibfnamefont {E.~Q.}\ \bibnamefont {Simmons}}, \bibinfo {author}
  {\bibfnamefont {F.~L.~N.}\ \bibnamefont {Santos}}, \bibinfo {author}
  {\bibfnamefont {P.}~\bibnamefont {Vignolo}}, \bibinfo {author} {\bibfnamefont
  {T.}~\bibnamefont {Macr\`{\i}}},\ and\ \bibinfo {author} {\bibfnamefont
  {D.~M.}\ \bibnamefont {Weld}},\ }\bibfield  {title} {\bibinfo {title}
  {Observation of the quantum boomerang effect},\ }\href
  {https://doi.org/10.1103/PhysRevX.12.011035} {\bibfield  {journal} {\bibinfo
  {journal} {Phys. Rev. X}\ }\textbf {\bibinfo {volume} {12}},\ \bibinfo
  {pages} {011035} (\bibinfo {year} {2022})}\BibitemShut {NoStop}%
\bibitem [{\citenamefont {Lin}\ \emph {et~al.}(2011)\citenamefont {Lin},
  \citenamefont {Jim{\'{e}}nez-Garc{\'{i}}a},\ and\ \citenamefont
  {Spielman}}]{Lin2011}%
  \BibitemOpen
  \bibfield  {author} {\bibinfo {author} {\bibfnamefont {Y.-J.}\ \bibnamefont
  {Lin}}, \bibinfo {author} {\bibfnamefont {K.}~\bibnamefont
  {Jim{\'{e}}nez-Garc{\'{i}}a}},\ and\ \bibinfo {author} {\bibfnamefont
  {I.~B.}\ \bibnamefont {Spielman}},\ }\bibfield  {title} {\bibinfo {title}
  {{Spin–orbit-coupled Bose–Einstein condensates}},\ }\href
  {https://doi.org/10.1038/nature09887} {\bibfield  {journal} {\bibinfo
  {journal} {Nature}\ }\textbf {\bibinfo {volume} {471}},\ \bibinfo {pages}
  {83} (\bibinfo {year} {2011})}\BibitemShut {NoStop}%
\bibitem [{\citenamefont {Hamner}\ \emph {et~al.}(2015)\citenamefont {Hamner},
  \citenamefont {Zhang}, \citenamefont {Khamehchi}, \citenamefont {Davis},\
  and\ \citenamefont {Engels}}]{Hamner2015}%
  \BibitemOpen
  \bibfield  {author} {\bibinfo {author} {\bibfnamefont {C.}~\bibnamefont
  {Hamner}}, \bibinfo {author} {\bibfnamefont {Y.}~\bibnamefont {Zhang}},
  \bibinfo {author} {\bibfnamefont {M.~A.}\ \bibnamefont {Khamehchi}}, \bibinfo
  {author} {\bibfnamefont {M.~J.}\ \bibnamefont {Davis}},\ and\ \bibinfo
  {author} {\bibfnamefont {P.}~\bibnamefont {Engels}},\ }\bibfield  {title}
  {\bibinfo {title} {{Spin-Orbit-Coupled Bose-Einstein Condensates in a
  One-Dimensional Optical Lattice}},\ }\href
  {https://doi.org/10.1103/PhysRevLett.114.070401} {\bibfield  {journal}
  {\bibinfo  {journal} {Phys. Rev. Lett.}\ }\textbf {\bibinfo {volume} {114}},\
  \bibinfo {pages} {070401} (\bibinfo {year} {2015})}\BibitemShut {NoStop}%
\bibitem [{\citenamefont {Yue}\ \emph {et~al.}(2020)\citenamefont {Yue},
  \citenamefont {{S{\'{a}} de Melo}},\ and\ \citenamefont
  {Spielman}}]{Yue2020a}%
  \BibitemOpen
  \bibfield  {author} {\bibinfo {author} {\bibfnamefont {Y.}~\bibnamefont
  {Yue}}, \bibinfo {author} {\bibfnamefont {C.~A.~R.}\ \bibnamefont {{S{\'{a}}
  de Melo}}},\ and\ \bibinfo {author} {\bibfnamefont {I.~B.}\ \bibnamefont
  {Spielman}},\ }\bibfield  {title} {\bibinfo {title} {{Enhanced transport of
  spin-orbit-coupled Bose gases in disordered potentials}},\ }\href
  {https://doi.org/10.1103/PhysRevA.102.033325} {\bibfield  {journal} {\bibinfo
   {journal} {Phys. Rev. A}\ }\textbf {\bibinfo {volume} {102}},\ \bibinfo
  {pages} {033325} (\bibinfo {year} {2020})}\BibitemShut {NoStop}%
\bibitem [{foo()}]{foot}%
  \BibitemOpen
  \href@noop {} {}\bibinfo {note} {In~\cite{Sajjad2021}, it was shown that the
  initial state also has to be time reversal invariant, a condition implicitly
  met by the initial states considered in~\cite{Prat2019} and in the present
  paper.}\BibitemShut {Stop}%
\bibitem [{\citenamefont {Haake}\ \emph {et~al.}(2018)\citenamefont {Haake},
  \citenamefont {Gnutzmann},\ and\ \citenamefont {Ku{\'{s}}}}]{Haake2018}%
  \BibitemOpen
  \bibfield  {author} {\bibinfo {author} {\bibfnamefont {F.}~\bibnamefont
  {Haake}}, \bibinfo {author} {\bibfnamefont {S.}~\bibnamefont {Gnutzmann}},\
  and\ \bibinfo {author} {\bibfnamefont {M.}~\bibnamefont {Ku{\'{s}}}},\ }\href
  {https://doi.org/10.1007/978-3-319-97580-1} {\emph {\bibinfo {title}
  {{Quantum Signatures of Chaos}}}},\ Springer Series in Synergetics\ (\bibinfo
   {publisher} {Springer International Publishing},\ \bibinfo {address}
  {Cham},\ \bibinfo {year} {2018})\BibitemShut {NoStop}%
\bibitem [{\citenamefont {Oganesyan}\ and\ \citenamefont
  {Huse}(2007)}]{Oganesyan2007}%
  \BibitemOpen
  \bibfield  {author} {\bibinfo {author} {\bibfnamefont {V.}~\bibnamefont
  {Oganesyan}}\ and\ \bibinfo {author} {\bibfnamefont {D.~A.}\ \bibnamefont
  {Huse}},\ }\bibfield  {title} {\bibinfo {title} {{Localization of interacting
  fermions at high temperature}},\ }\href
  {https://doi.org/10.1103/PhysRevB.75.155111} {\bibfield  {journal} {\bibinfo
  {journal} {Phys. Rev. B}\ }\textbf {\bibinfo {volume} {75}},\ \bibinfo
  {pages} {155111} (\bibinfo {year} {2007})}\BibitemShut {NoStop}%
\bibitem [{\citenamefont {Atas}\ \emph {et~al.}(2013)\citenamefont {Atas},
  \citenamefont {Bogomolny}, \citenamefont {Giraud},\ and\ \citenamefont
  {Roux}}]{Atas2013a}%
  \BibitemOpen
  \bibfield  {author} {\bibinfo {author} {\bibfnamefont {Y.~Y.}\ \bibnamefont
  {Atas}}, \bibinfo {author} {\bibfnamefont {E.}~\bibnamefont {Bogomolny}},
  \bibinfo {author} {\bibfnamefont {O.}~\bibnamefont {Giraud}},\ and\ \bibinfo
  {author} {\bibfnamefont {G.}~\bibnamefont {Roux}},\ }\bibfield  {title}
  {\bibinfo {title} {{Distribution of the ratio of consecutive level spacings
  in random matrix ensembles}},\ }\href
  {https://doi.org/10.1103/PhysRevLett.110.084101} {\bibfield  {journal}
  {\bibinfo  {journal} {Phys. Rev. Lett.}\ }\textbf {\bibinfo {volume} {110}},\
  \bibinfo {pages} {1} (\bibinfo {year} {2013})}\BibitemShut {NoStop}%
\bibitem [{\citenamefont {Prat}(2017)}]{Prat2017a}%
  \BibitemOpen
  \bibfield  {author} {\bibinfo {author} {\bibfnamefont {T.}~\bibnamefont
  {Prat}},\ }\emph {\bibinfo {title} {{Anderson localization with cold atoms :
  dynamics in disorder and prospects from chaos}}},\ \href
  {https://tel.archives-ouvertes.fr/tel-01687032} {Ph.D. thesis},\ \bibinfo
  {school} {Universit{\'{e}} Pierre et Marie Curie (Paris)} (\bibinfo {year}
  {2017})\BibitemShut {NoStop}%
\bibitem [{\citenamefont {Akkermans}\ and\ \citenamefont
  {Montambaux}(2007)}]{Akkermans2007}%
  \BibitemOpen
  \bibfield  {author} {\bibinfo {author} {\bibfnamefont {E.}~\bibnamefont
  {Akkermans}}\ and\ \bibinfo {author} {\bibfnamefont {G.}~\bibnamefont
  {Montambaux}},\ }\href {https://doi.org/10.1017/CBO9780511618833} {\emph
  {\bibinfo {title} {Mesoscopic Phys. Electrons Photons}}},\ Vol.\ \bibinfo
  {volume} {9780521855}\ (\bibinfo  {publisher} {Cambridge university press},\
  \bibinfo {year} {2007})\ pp.\ \bibinfo {pages} {1--588}\BibitemShut {NoStop}%
\bibitem [{\citenamefont {M{\"{u}}ller}\ and\ \citenamefont
  {Delande}(2011)}]{Muller2011}%
  \BibitemOpen
  \bibfield  {author} {\bibinfo {author} {\bibfnamefont {C.~A.}\ \bibnamefont
  {M{\"{u}}ller}}\ and\ \bibinfo {author} {\bibfnamefont {D.}~\bibnamefont
  {Delande}},\ }\bibfield  {title} {\bibinfo {title} {{Disorder and
  interference: localization phenomena}},\ }in\ \href
  {https://doi.org/10.1093/acprof:oso/9780199603657.003.0009} {\emph {\bibinfo
  {booktitle} {Ultracold Gases Quantum Inf.}}}\ (\bibinfo  {publisher} {Oxford
  University Press},\ \bibinfo {year} {2011})\ pp.\ \bibinfo {pages}
  {441--533}\BibitemShut {NoStop}%
\bibitem [{sup()}]{supple}%
  \BibitemOpen
  \href@noop {} {}\bibinfo {note} {See Supplemental Material, that includes
  Ref.~\cite{Abramowitz1966}, at [URL will be inserted by publisher], for the
  solution of the classical coupled Boltzmann equations, a sketch of the
  diagrammatic method used to compute the quantum boomerang effect in the
  absence of time-reversal invariance, the contributions of the various
  components to the center of mass position, and the dependence on the TRI
  breaking parameter $\Delta$}\BibitemShut {NoStop}%
\bibitem [{\citenamefont {Tal-Ezer}\ and\ \citenamefont
  {Kosloff}(1984)}]{Tal-Ezer1984}%
  \BibitemOpen
  \bibfield  {author} {\bibinfo {author} {\bibfnamefont {H.}~\bibnamefont
  {Tal-Ezer}}\ and\ \bibinfo {author} {\bibfnamefont {R.}~\bibnamefont
  {Kosloff}},\ }\bibfield  {title} {\bibinfo {title} {{An accurate and
  efficient scheme for propagating the time dependent Schr{\"{o}}dinger
  equation}},\ }\href {https://doi.org/10.1063/1.448136} {\bibfield  {journal}
  {\bibinfo  {journal} {J. Chem. Phys.}\ }\textbf {\bibinfo {volume} {81}},\
  \bibinfo {pages} {3967} (\bibinfo {year} {1984})}\BibitemShut {NoStop}%
\bibitem [{\citenamefont {Leforestier}\ \emph {et~al.}(1991)\citenamefont
  {Leforestier}, \citenamefont {Bisseling}, \citenamefont {Cerjan},
  \citenamefont {Feit}, \citenamefont {Friesner}, \citenamefont {Guldberg},
  \citenamefont {Hammerich}, \citenamefont {Jolicard}, \citenamefont
  {Karrlein}, \citenamefont {Meyer}, \citenamefont {Lipkin}, \citenamefont
  {Roncero},\ and\ \citenamefont {Kosloff}}]{Leforestier1991}%
  \BibitemOpen
  \bibfield  {author} {\bibinfo {author} {\bibfnamefont {C.}~\bibnamefont
  {Leforestier}}, \bibinfo {author} {\bibfnamefont {R.}~\bibnamefont
  {Bisseling}}, \bibinfo {author} {\bibfnamefont {C.}~\bibnamefont {Cerjan}},
  \bibinfo {author} {\bibfnamefont {M.}~\bibnamefont {Feit}}, \bibinfo {author}
  {\bibfnamefont {R.}~\bibnamefont {Friesner}}, \bibinfo {author}
  {\bibfnamefont {A.}~\bibnamefont {Guldberg}}, \bibinfo {author}
  {\bibfnamefont {A.}~\bibnamefont {Hammerich}}, \bibinfo {author}
  {\bibfnamefont {G.}~\bibnamefont {Jolicard}}, \bibinfo {author}
  {\bibfnamefont {W.}~\bibnamefont {Karrlein}}, \bibinfo {author}
  {\bibfnamefont {H.-D.}\ \bibnamefont {Meyer}}, \bibinfo {author}
  {\bibfnamefont {N.}~\bibnamefont {Lipkin}}, \bibinfo {author} {\bibfnamefont
  {O.}~\bibnamefont {Roncero}},\ and\ \bibinfo {author} {\bibfnamefont
  {R.}~\bibnamefont {Kosloff}},\ }\bibfield  {title} {\bibinfo {title} {{A
  comparison of different propagation schemes for the time dependent
  Schr{\"{o}}dinger equation}},\ }\href
  {https://doi.org/10.1016/0021-9991(91)90137-A} {\bibfield  {journal}
  {\bibinfo  {journal} {J. Comput. Phys.}\ }\textbf {\bibinfo {volume} {94}},\
  \bibinfo {pages} {59} (\bibinfo {year} {1991})}\BibitemShut {NoStop}%
\bibitem [{\citenamefont {Roche}\ and\ \citenamefont
  {Mayou}(1997)}]{Roche1997}%
  \BibitemOpen
  \bibfield  {author} {\bibinfo {author} {\bibfnamefont {S.}~\bibnamefont
  {Roche}}\ and\ \bibinfo {author} {\bibfnamefont {D.}~\bibnamefont {Mayou}},\
  }\bibfield  {title} {\bibinfo {title} {{Conductivity of Quasiperiodic
  Systems: A Numerical Study}},\ }\href
  {https://doi.org/10.1103/PhysRevLett.79.2518} {\bibfield  {journal} {\bibinfo
   {journal} {Phys. Rev. Lett.}\ }\textbf {\bibinfo {volume} {79}},\ \bibinfo
  {pages} {2518} (\bibinfo {year} {1997})}\BibitemShut {NoStop}%
\bibitem [{\citenamefont {Wei{\ss}e}\ \emph {et~al.}(2006)\citenamefont
  {Wei{\ss}e}, \citenamefont {Wellein}, \citenamefont {Alvermann},\ and\
  \citenamefont {Fehske}}]{Weisse2006}%
  \BibitemOpen
  \bibfield  {author} {\bibinfo {author} {\bibfnamefont {A.}~\bibnamefont
  {Wei{\ss}e}}, \bibinfo {author} {\bibfnamefont {G.}~\bibnamefont {Wellein}},
  \bibinfo {author} {\bibfnamefont {A.}~\bibnamefont {Alvermann}},\ and\
  \bibinfo {author} {\bibfnamefont {H.}~\bibnamefont {Fehske}},\ }\bibfield
  {title} {\bibinfo {title} {{The kernel polynomial method}},\ }\href
  {https://doi.org/10.1103/RevModPhys.78.275} {\bibfield  {journal} {\bibinfo
  {journal} {Rev. Mod. Phys.}\ }\textbf {\bibinfo {volume} {78}},\ \bibinfo
  {pages} {275} (\bibinfo {year} {2006})}\BibitemShut {NoStop}%
\bibitem [{\citenamefont {Janarek}\ \emph {et~al.}(2020)\citenamefont
  {Janarek}, \citenamefont {Delande}, \citenamefont {Cherroret},\ and\
  \citenamefont {Zakrzewski}}]{Janarek2020a}%
  \BibitemOpen
  \bibfield  {author} {\bibinfo {author} {\bibfnamefont {J.}~\bibnamefont
  {Janarek}}, \bibinfo {author} {\bibfnamefont {D.}~\bibnamefont {Delande}},
  \bibinfo {author} {\bibfnamefont {N.}~\bibnamefont {Cherroret}},\ and\
  \bibinfo {author} {\bibfnamefont {J.}~\bibnamefont {Zakrzewski}},\ }\bibfield
   {title} {\bibinfo {title} {{Quantum boomerang effect for interacting
  particles}},\ }\href {https://doi.org/10.1103/PhysRevA.102.013303} {\bibfield
   {journal} {\bibinfo  {journal} {Phys. Rev. A}\ }\textbf {\bibinfo {volume}
  {102}},\ \bibinfo {pages} {1} (\bibinfo {year} {2020})}\BibitemShut {NoStop}%
\bibitem [{\citenamefont {Berezinskiĭ}(1973)}]{Berezinskii1996}%
  \BibitemOpen
  \bibfield  {author} {\bibinfo {author} {\bibfnamefont {V.~L.}\ \bibnamefont
  {Berezinskiĭ}},\ }\bibfield  {title} {\bibinfo {title} {{Kinetics of a
  quantum particle in a one-dimensional random potential [ Kinetika kvantovoi
  chastitsy v odnomernom sluchainom potentsiale]}},\ }\href
  {http://www.jetp.ac.ru/cgi-bin/index/e/38/3/p620?a=list} {\bibfield
  {journal} {\bibinfo  {journal} {Zhurnal Eksp. i Teor. Fiz.}\ }\textbf
  {\bibinfo {volume} {65}},\ \bibinfo {pages} {1251} (\bibinfo {year}
  {1973})}\BibitemShut {NoStop}%
\bibitem [{\citenamefont {Gogolin}(1976)}]{Gogolin1976}%
  \BibitemOpen
  \bibfield  {author} {\bibinfo {author} {\bibfnamefont {A.}~\bibnamefont
  {Gogolin}},\ }\bibfield  {title} {\bibinfo {title} {{Electron density
  distribution for localized states in a one-dimensional disordered system}},\
  }\href@noop {} {\bibfield  {journal} {\bibinfo  {journal} {Sov. J. Exp.
  Theor. Phys.}\ }\textbf {\bibinfo {volume} {44}},\ \bibinfo {pages} {1003}
  (\bibinfo {year} {1976})}\BibitemShut {NoStop}%
\bibitem [{\citenamefont {Nakhmedov}\ \emph {et~al.}(1987)\citenamefont
  {Nakhmedov}, \citenamefont {Prigodin},\ and\ \citenamefont
  {Firsov}}]{Nakhmedov1987}%
  \BibitemOpen
  \bibfield  {author} {\bibinfo {author} {\bibfnamefont {E.}~\bibnamefont
  {Nakhmedov}}, \bibinfo {author} {\bibfnamefont {V.}~\bibnamefont
  {Prigodin}},\ and\ \bibinfo {author} {\bibfnamefont {Y.}~\bibnamefont
  {Firsov}},\ }\bibfield  {title} {\bibinfo {title} {{Localization dynamics in
  weakly disordered systems}},\ }\href@noop {} {\bibfield  {journal} {\bibinfo
  {journal} {Zhurnal Eksp. i Teor. Fiz.}\ }\textbf {\bibinfo {volume} {92}},\
  \bibinfo {pages} {2133} (\bibinfo {year} {1987})}\BibitemShut {NoStop}%
\bibitem [{\citenamefont {Valdes}\ and\ \citenamefont
  {Wellens}(2016)}]{Wellens2016}%
  \BibitemOpen
  \bibfield  {author} {\bibinfo {author} {\bibfnamefont {J.~P.~R.}\
  \bibnamefont {Valdes}}\ and\ \bibinfo {author} {\bibfnamefont
  {T.}~\bibnamefont {Wellens}},\ }\bibfield  {title} {\bibinfo {title}
  {{Localization of wave packets in one-dimensional random potentials}},\
  }\href {https://doi.org/10.1103/PhysRevA.93.063634} {\bibfield  {journal}
  {\bibinfo  {journal} {Phys. Rev. A}\ }\textbf {\bibinfo {volume} {93}},\
  \bibinfo {pages} {063634} (\bibinfo {year} {2016})}\BibitemShut {NoStop}%
\bibitem [{\citenamefont {Janarek}\ \emph {et~al.}()\citenamefont {Janarek},
  \citenamefont {Prat}, \citenamefont {Delande},\ and\ \citenamefont
  {Cherroret}}]{Berezinskii_method:to_be_published}%
  \BibitemOpen
  \bibfield  {author} {\bibinfo {author} {\bibfnamefont {J.}~\bibnamefont
  {Janarek}}, \bibinfo {author} {\bibfnamefont {T.}~\bibnamefont {Prat}},
  \bibinfo {author} {\bibfnamefont {D.}~\bibnamefont {Delande}},\ and\ \bibinfo
  {author} {\bibfnamefont {N.}~\bibnamefont {Cherroret}},\ }\bibinfo {note} {in
  preparation}\BibitemShut {NoStop}%
\bibitem [{\citenamefont {Baker}(1975)}]{Baker1975}%
  \BibitemOpen
  \bibfield  {author} {\bibinfo {author} {\bibfnamefont {G.~A.}\ \bibnamefont
  {Baker}},\ }\href
  {https://www.elsevier.com/books/essentials-of-pade-approximants/baker/978-0-12-074855-6}
  {\emph {\bibinfo {title} {{Essentials of Pad{\'{e}} approximants}}}}\
  (\bibinfo  {publisher} {Academic Press},\ \bibinfo {year} {1975})\BibitemShut
  {NoStop}%
\bibitem [{\citenamefont {Noronha}\ and\ \citenamefont
  {Macrì}(2022)}]{Macri22}%
  \BibitemOpen
  \bibfield  {author} {\bibinfo {author} {\bibfnamefont {F.}~\bibnamefont
  {Noronha}}\ and\ \bibinfo {author} {\bibfnamefont {T.}~\bibnamefont
  {Macrì}},\ }\bibfield  {title} {\bibinfo {title} {Ubiquity of the quantum
  boomerang effect in localized systems},\ }\href
  {https://doi.org/10.48550/ARXIV.2203.16719} {\bibfield  {journal} {\bibinfo
  {journal} {arXiv e-prints:}\ ,\ \bibinfo {pages} {arXiv:2203.16719}}
  (\bibinfo {year} {2022})}\BibitemShut {NoStop}%
\bibitem [{\citenamefont {Abramowitz}\ and\ \citenamefont
  {Stegun}(1966)}]{Abramowitz1966}%
  \BibitemOpen
  \bibfield  {author} {\bibinfo {author} {\bibfnamefont {M.}~\bibnamefont
  {Abramowitz}}\ and\ \bibinfo {author} {\bibfnamefont {I.~A.}\ \bibnamefont
  {Stegun}},\ }\href@noop {} {\emph {\bibinfo {title} {{Handbook of
  Mathematical Functions}}}}\ (\bibinfo  {publisher} {National Bureau of
  Standards},\ \bibinfo {year} {1966})\BibitemShut {NoStop}%
\end{thebibliography}
\end{document}